\documentclass[aps,twocolumn,eqsecnum,showpacs]{revtex4}
\usepackage{amsthm}
\usepackage{amsmath}
\usepackage{latexsym}
\usepackage{amsfonts}
\usepackage{amssymb}
\usepackage{color}
\usepackage{bbm,dsfont}
\usepackage{graphicx}

\theoremstyle{definition}

\newtheorem{definition}{Definition}

\newcommand{\ip}[2]{\left\langle\,#1\,|\,#2\,\right\rangle} 
\newcommand{\ket}[1]{|#1\rangle} 
\newcommand{\bra}[1]{\langle#1|} 
\newcommand{\tr}[1]{\textrm{tr}\left[#1\right]} 

\newcommand{\A}{\mathcal{A}}
\newcommand{\B}{\mathcal{B}}
\newcommand{\E}{\mathcal{E}}
\newcommand{\M}{\mathcal{M}}

\newcommand{\R}{\mathcal{R}}%

\newcommand{\cO}{{\mathcal{O}}}

\newcommand{\cH}{{\mathcal{H}}}

\newcommand{\be}{\begin{eqnarray}}
\newcommand{\ee}{\end{eqnarray}}

\begin{document}

\title{Unambiguous comparison of quantum measurements}
\author{Mario Ziman$^{1,2}$, Teiko Heinosaari$^{3}$ and Michal Sedl\'ak$^{1}$}
\affiliation{$^{1}$Institute of Physics, Slovak Academy of Sciences, D\'ubravsk\'a cesta 9, 845 11 Bratislava, Slovakia\\
$^{2}$Faculty of Informatics, Masaryk University, Botanick\'a 68a, Brno, Czech Republic\\
$^{3}$Niels Bohr Institute, Blegdamsvej 17, 2100 Copenhagen, Denmark
}
\date{\today}

\begin{abstract}
The goal of comparison is to reveal the difference of compared objects as
fast and reliably as possible. In this paper we formulate and investigate
the unambiguous comparison of
unknown quantum measurements represented by non-degenerate sharp
POVMs.
We distinguish between measurement devices with apriori labeled and
unlabeled outcomes. In both cases we can unambiguously conclude only that the
measurements are  different. For the labeled case it is sufficient
to use each unknown measurement only once and the average conditional
success probability decreases with the Hilbert space dimension as $1/d$. If the
outcomes of the apparatuses are not labeled, then the problem is more complicated.
We analyze the case of two-dimensional Hilbert space. In this case
single shot comparison is impossible and each measurement
device must be used (at least) twice. The optimal test state in the two-shots
scenario gives the average conditional success probability $3/4$. Interestingly,
the optimal experiment detects unambiguously the difference with
nonvanishing probability for any pair of observables.
\end{abstract}
\maketitle

\section{Introduction}\label{sec:intro}

Unavoidable randomness is one of the most important paradigms
of quantum theory. As a consequence, its typical predictions and conclusions
have a statistical and probabilistic essence. However, there are exceptions \cite{Chefles00,Ivanovic87,Ji06,Ziman08}.
For example, a photon passing through a vertical polarizer
will pass the second vertical polarizer with probability 1. In such ``certain'' cases, the goal is not to acquire a complete description of quantum systems, but rather to identify some features of interest.
In this paper we investigate a variant of an unambiguous quantum
comparison problem \cite{Barnett03,Andersson03,Andersson06,Sedlak09},
i.e., a task in which the aim is to compare a pair of quantum devices.

It is an interesting question how the quantum systems can be compared
and which of the quantum properties are comparable. For example, the velocity of quantum particles
is a well defined property only under very specific conditions.
In general, the probability distribution of velocities is the correct
description of the dynamical properties of quantum particles. Therefore,
in quantum case it is typical that the comparison problem is naturally
a statistical problem. This is in contradiction with the usual approach
to comparison tasks, which are based on individual events rather than on
statistics. It could seem counter-intuitive,
but individual experimental clicks can provide us with a definite and
unambiguous answer even if the description is statistical. In general,
the goal is to design an experiment accepting quantum devices
as free parameters and producing events we can associate with three
conclusions: i) same, ii) different, and iii) no conclusion.

So far, the unambiguous comparison problem has been studied in the cases
of pure states \cite{Barnett03,Andersson06}
and unitary channels \cite{Andersson03,Sedlak09}.
In this paper we analyze the unambiguous comparison of quantum measurements.
Suppose that we are
given a pair of experimental setups implementing qubit measurements,
each of them designed by a different experimentalist. Is there
a way to unambiguously compare their performance? Especially, are they same or different? As independent
experimentalists we can think of these experimental setups as black boxes,
producing outcomes after a qubit is inserted. Our conclusions then have to be based on the acquired measurement outcomes.

For quantum measurements, there are two natural variations
of the comparison problem.
First of all, we can ask whether the given black boxes are identical.
This means that they produce the same measurement outcome statistics in any state. In particular, also the labeling of the outcomes is similar. For instance, two Stern-Gerlach apparatuses oriented in opposite directions are considered to be different in this strict sense. However, they can be made identical by simply re-labeling the outcomes in one of them. Thus, the other way to compare two black boxes is to ask whether they are {\em equivalent}, i.e.,  identical after suitable re-labeling of the outcomes.

As an example, suppose we are comparing whether two Stern-Gerlach apparatuses are identical.
A singlet state of two qubits inserted into the measurements cannot lead to the same outcomes
unless the measurement devices (including the labeling) are different.
If labeling of the outcomes is not given or it is part of the comparison problem, then we can perform this singlet-based test for all possible labelings independently. Finding the same
unambiguous conclusions in all of them leads to a conclusion
also for measurements without apriori labels. Since for each of the
Stern-Gerlach apparatuses we have two different choices of labels,
we need to perform the singlet-based comparison four times,
i.e. each of the apparatuses is used 4 times. We will show that there are also better
 strategies in which each of the unlabeled apparatuses is used only twice.

The paper is organized as follows. In Section \ref{sec:observables} we shortly
recall the mathematical description of quantum observables. Sections
\ref{sec:black_box} and \ref{sec:apriori} explain the concepts of unknown quantum measurement apparatuses
and apriori information. The unambiguous comparison of measurements
with labeled outcomes is presented in Section \ref{sec:labeled_comparison} and for apparatuses
with unlabeled outcomes in Section \ref{sec:unlabeled}. In the last Section \ref{sec:summary} we summarize
the obtained results. Some of the techical details are given in the Appendix.

\section{Observables}\label{sec:observables}

The statistics of quantum measurements is described
by positive operator valued measures (POVM). In what follows
we consider only measurements with finite number of
outcomes. For simplicity, we assume that these
outcomes form an index set $J_n=\{1,\dots,n\}$.
The associated POVM is a mapping $\A$ from $J_n$ into
the set of effects $\E(\cH)$, i.e. a set of positive operators $E$
on Hilbert space $\cH$ such that $O\leq E\leq I$, where $O$ is the zero
operator and $I$ is the identity operator. Moreover, the POVM is
normalized to identity i.e. $\A_1+\cdots+\A_n=I$,
where $\A_i\equiv \A(i)$.
The effects serve
as a proper mathematical representation of observed
quantum events (experimental clicks). A probability to observe an effect $E$
is given by the trace formula
\be
p_E=\tr{\varrho E}\, ,
\ee
where $\varrho$ is a state of the measured quantum system.

For an operator $X$, we denote by $\Pi_X$ the projection
onto the support of $X$. For effects and states we then have $E\leq\Pi_E$ and
$\varrho\leq\Pi_\varrho$. Moreover, the condition $\tr{\varrho E}>0$ is equivalent to
$\Pi_E\Pi_\varrho\neq O$.

We say that observable is {\it sharp} if each effect composing the POVM
is a projection, i.e., $E_j=E_j^2$ for all $j$. If, moreover,
$E_j\cH$ is a one-dimensional subspace of $\cH$ for each $j$,
then the observable is {\it non-degenerate}. In such case
we can write $E_j=\ket{\psi_j}\bra{\psi_j}\equiv\psi_j$ and
$\ip{\psi_j}{\psi_k}=\delta_{jk}$. In fact, each orthonormal basis of the Hilbert space defines a sharp non-degenerate POVM. We denote by $\M$ the set of
all non-degenerate sharp observables. It is closed
under the action of the unitary group $U(d)$ transforming
$\A=\{\A_j\}\in \M$ into an observable $\A^U$ consisting of
effects $\A^U_{j}=U\A_j U^\dagger$.

\section{Unknown black box}\label{sec:black_box}

We shall think of an unknown measurement apparatus as of a black box accepting
physical systems and producing one of $n$ distinguished outcomes.
For sharp non-degenerate observables
each of the outcomes is associated with a one-dimensional
projection. We distinguish two types of black boxes leading to
two different concepts of equivalence of observables and affecting the
formulation of the comparison problem, too. In principle,
we can meet with measurement outcomes that are
either labeled, or not. If the outcomes are not labeled, we
assign a number $j\in J_n$ to each of them. However, in such case
the ambiguity of relabeling must be taken into account and equivalence
of observables should be compatible with this freedom. Let us spell these definitions explicitly.

\begin{definition}
Observables $\A:J_n\to\E(\cH)$ and $\B:J_n\to\E(\cH)$
are \emph{identical} if $\A_j=\B_j$ for all $j$.
\end{definition}

\begin{definition}
Observables $\A:J_n\to\E(\cH)$ and $\B:J_n\to\E(\cH)$
are \emph{equivalent} (in the unlabeled sense) if there exist a
permutation $\pi: J_n\to J_n$ such that $\A_j=\B_{\pi(j)}$ for all $j$.
\end{definition}

It follows from the definition that equivalence
class of an unlabeled observable consists of POVMs
with the same range, i.e. the elements of the set
of unlabeled measurements can be understood as unordered
collections of effects summing up to identity.
The comparison of unlabeled measurements can hence be seen as a comparison
of ranges of POVMs.

A single usage of a measurement device tells us
that an effect $E$ associated with the observed outcome has support
overlapping with the support of $\varrho$, i.e. $\Pi_E\Pi_\varrho\neq O$.
However, in the unlabeled case this information does not tell us
too much about the particular effect associated with the observed
outcome. Let us consider an unlabeled measurement described by
effects $\{\A_1,\dots,\A_n\}$ forming a particular POVM once the
ordering is fixed. In fact, since we assume that the
labeling is chosen in random way, for each artificially named outcome
the predicted probability is the same, i.e.
\be
\nonumber
p_j(\A)= \frac{1}{n!}\sum_\pi \tr{\varrho \A_{\pi(j)}}
=\frac{(n-1)!}{n!}\sum_{j^\prime} \tr{\varrho\A_{j^\prime}}=\frac{1}{n}\, ,
\ee
where we used the fact that $n!$ is the total number of permutations
on $J_n$, $(n-1)!$ is the number of them having a specific
label $j^\prime$ on the fixed ($j$th) position.

Using the apparatus once more we can distinguish whether
the observed outcomes coincide, or not. After fixing the labels
$1,\dots,n$ of the measurement device, the probability to observe
a pair of outcomes $j,k$ reads
\be
\label{eq:prva}
p_{jk}(\A)&=&\frac{1}{n!}\sum_\pi \tr{\varrho \A_{\pi(j)}\otimes \A_{\pi(k)}}\\
\nonumber
&=& \frac{(n-1)!}{n!}\sum_{j^\prime}\tr{\varrho\A_{j^\prime}\otimes\A_{j^\prime}}\qquad {\rm if}\ j=k\,;\\
\nonumber
&=& \frac{(n-2)!}{n!}\sum_{j^\prime\neq k^\prime}\tr{\varrho\A_{j^\prime}\otimes\A_{k^\prime}}\qquad {\rm if}\ j\neq k\,,
\ee
where $(n-2)!$ is the number of permutations resulting in fixed operators $\A_{j^\prime},\A_{k^\prime}$ for outcomes $j,k$.
Let us note that the values of $p_{jk}$ do not depend on particular
values of $j,k$, but only on their relative relation whether
$j=k$, or $j\neq k$. Consequently,
the probability to find the same/different outcomes in two shots
reads
\be
\nonumber
p_{\rm same}&=&n p_{jj}=
\sum_j\tr{\varrho\A_j\otimes\A_j}\, ,\\
\nonumber
p_{\rm diff}&=&n(n-1)p_{jk}=\sum_{j\neq k}\tr{\varrho\A_j\otimes\A_k}\, .
\ee
We used the fact that for $n$-valued measurement used twice
there are in total $n$ pairs of same outcomes and $n(n-1)$ pairs
of different outcomes. In this two-shot scenario the probabilities
$p_{\rm same},p_{\rm diff}$ depend on particular properties of effects
$\A_1,\dots,\A_n$, hence they contain some information
about $\A$.

\section{Apriori information}\label{sec:apriori}

From now on, we assume that otherwise unknown measurement apparatuses
are described by sharp non-degenerate observables. This assumption represents
a very important part of our apriori information. As such, they are
in direct correspondence
with orthonormal bases and have the same number of outcomes as the dimension of the Hilbert space ($n=d$).  Let us fix an orthonormal basis
$\ket{\psi_1},\dots,\ket{\psi_d}$
and denote by $\A_j^U$ the projections onto vectors $U\ket{\psi_j}$,
where $U$ is a unitary operator defined on $\cH$.
The projections $\A_1^U,\dots,\A_d^U$ form a non-degenerate sharp
observable $\A^U$. Moreover, every non-degenerate sharp observable is of the form $\A^U$
 for some unitary operator $U$.  We assume that each $\A^U$ is equally likely, therefore we have to average our expectations over all observables $\A^U$ using the Haar measure on the unitary group $U(d)$.

If the outcomes are labeled, then due to our apriori information
a particular sequence of outcomes
$\vec{j}=(j_1,\dots,j_r)\in J_d\times\cdots\times J_d$ can be observed
in $r$ usages of the apparatus with the average probability
\be
\overline{q}_{\vec{j}}=\int dU \tr{\varrho \A_{j_1}^U
\otimes\cdots\otimes \A_{j_r}^U}\, .
\ee

Further, let us discuss how the considered apriori information
affects the formulas for probabilities in the unlabeled case. For
the purposes of later analysis it is sufficient to investigate
only the experiments in which the apparatus is used at most twice.
 Thus, if the observable is unlabeled and $r=2$, then
the average probability to observe the outcomes $j,k$ reads
\be
\overline{p}_{jk}=\int dU p_{jk}(\A^U)\, ,
\ee
where $p_{jk}(\A^U)$ is specified in Eq.\eqref{eq:prva}.
Since
\be
\int dU \A^U_j\otimes\A^U_k =\int dU \A^U_{j^\prime} \otimes \A_{k^\prime}^U
\ee
for all $j\neq k,j^\prime\neq k^\prime$ and $j=k,j^\prime=k^\prime$,
respectively, it follows that
\be
\nonumber
\sum_j\int dU \A_j^U\otimes \A_j^U&=&d\int d\psi\  \psi\otimes\psi\,,\\
\nonumber
\sum_{j\neq k}\int dU \A_j^U\otimes \A_k^U&=&d(d-1)\int d\psi d\psi_\perp\
\psi\otimes\psi_\perp\,,
\ee
where $d\psi_\perp$ denotes the integration over all vectors orthogonal to $\psi$.
To simplify the expressions we replaced the integration over unitary
group by integration over pure states $\psi$. In summary, we get
\be
\overline{p}_{jj}&=&\int d\psi \ \tr{\varrho\psi\otimes\psi}\,,\\
\overline{p}_{jk}&=&\int d\psi d\psi_\perp \ \tr{\varrho\psi\otimes\psi_\perp}\,.
\ee

We see that $\overline{p}_{jk}$ and $\overline{p}_{jj}$ do not
depend on particular values of indexes $j,k$, which are anyway
chosen by us and cannot be distinguished. As before,
we can discriminate only whether the outcomes are the
same, or different, with probabilities
given by formulas
\be
\nonumber
p_{\rm same}&=&d \overline{p}_{jj}=d\int d\psi \ \tr{\varrho\psi\otimes\psi}\,,\\
\nonumber
p_{\rm diff}&=&d(d-1)\overline{p}_{jk}=d(d-1)\int d\psi d\psi_\perp \  \tr{\varrho\psi\otimes\psi_\perp}\,.
\ee

In comparison, for labeled
observables in two shots we distinguish $d^2$ different outcomes
with probabilities
\be
\nonumber
\overline{q}_{jj}=\int  d\psi \ \tr{\varrho \psi\otimes\psi}\,,\quad
\overline{q}_{jk}=\int d\psi d\psi_\perp \ \tr{\varrho \psi\otimes\psi_\perp}\,.
\ee

\section{Comparison of labeled observables}\label{sec:labeled_comparison}

In the considered measurement comparison problem we are given
a pair of measurement devices measuring some non-degenerate
sharp observable $\A$ and $\B$. In this section we assume that
the outcomes of these devices are labeled by numbers $1,\dots,d$.
We start with the simplest experimental scenario
in which each of the apparatuses is used only once. Our goal is to find
a test state $\varrho$ and divide the potential outcomes $(j,k)$ into
three families associated with three conclusions: i) observables are identical,
 ii) observables are different (not identical), iii) no conclusion (inconclusive result).

Using a pair of labeled measurements (each of them once)
we distinguish $d^2$ different outcomes $(j,k)$ appearing
with probabilities $q_{jk}$ that depend on the equivalence of
$\A$ and $\B$
\be
\overline{q}_{jk}(\A\neq\B)&=&\int dU dV \tr{\varrho \A^U_j\otimes\B_k^V}\,,\\
\overline{q}_{jk}(\A=\B)&=&\int dU \tr{\varrho \A^U_j\otimes\A_k^U}\,.
\ee
Our a prior information manifested in the integration $\int dU$ causes that
probabilities $p_{jk}(\A\neq\B)$ and $p_{jk}(\A=\B)$
do not depend on particular values of $j,k$, but only
on their mutual relation $j=k$, or $j\neq k$.
That is, whatever test state is used, we can split the outcomes
at most into two classes, hence only two out of three conclusions
can be made.

In general, conclusion $y$ based on the observation of an outcome $x$
is unambiguous, if for all possible options except $y$
the conditional probability $p_x(z\neq y)$ vanishes. Since in our case the outcomes $(j,k)$
are divided into two subsets, $x\in\{\rm same,diff\}$, in order to
conclude that the observables are different the condition
$\overline{q}_x(\A=\B)=0$ must hold for some outcome $x$.
Similarly, if we can conclude that $\A=\B$, then
there must exist an outcome $x$ such that $\overline{q}_x(\A\neq\B)=0$.
We refer to such conditions as the \emph{unambiguous
no-error conditions}. Their validity is necessary in order to
call formulation and solution of the problem unambiguous. Outcomes not associated
with unambiguous conclusions lead to an inconclusive result. The smaller
is the probability of the inconclusive outcome the better is the
solution.

Let us note that
\be
\int d\psi\, \psi^{\otimes k}= \frac{(d-1)!k!}{(d+k-1)!}P^+_{1\dots k}\,,
\ee
where $P^+_{1\dots k}$ is the projection onto the
completely symmetric subspace of
$\cH^{\otimes k}$ and $d_k=\tr{P_{1\dots k}^+}=\frac{(d+k-1)!}{(d-1)!k!}$
the dimension of that subspace. For a fixed vector $\psi$
\be
\nonumber
\int d\psi_\perp\, \psi_\perp^{\otimes k}&=&\int_{\cH_\psi^\perp} d\varphi\,
\varphi^{\otimes k} \\&=&
\frac{(d-2)!k!}{(d+k-2)!}
(I-\psi)^{\otimes k}P^+_{1\dots k}\,,
\ee
where we used $\cH_\psi^\perp$ to denote the subspace of $\cH$ orthogonal
to $\ket{\psi}\in\cH$.

We will use these identities in the evaluation of the probabilities
$p_{jj}$ and $p_{jk}$. In particular,
\be
\overline{q}_{jj}(\A\neq\B)&=&\int d\psi d\varphi\,  \tr{\varrho\psi\otimes\varphi}=\frac{1}{d^2}\tr{\varrho}\,,\\
\overline{q}_{jk}(\A\neq\B)&=&\int d\psi d\varphi\,  \tr{\varrho\psi\otimes\varphi}=
\frac{1}{d^2}\tr{\varrho}\,,\\
\overline{q}_{jj}(\A=\B)&=&\int d\psi\,  \tr{\varrho\psi\otimes\psi}=\frac{1}{d_2}\tr{\varrho P_{12}^+}\,,\\
\nonumber
\overline{q}_{jk}(\A=\B)&=&\int d\psi\, \tr{\varrho\psi\otimes\psi_\perp}\\
 &=&
\frac{1}{d-1}\tr{\varrho(\frac{1}{d}I\otimes I-\frac{1}{d_2}P_{12}^+)}\, .
\ee

We see that if the measurement devices are different ($\A\neq\B$),
then for all test states $\varrho$ the probabilities
$\overline{q}_{jj}(\A\neq\B)$ and $\overline{q}_{jk}(\A\neq\B)$
do not vanish for any outcome. Because of that
the equality of the observables cannot be concluded unambiguously.

Denoting by $P_{12}^-=I\otimes I-P_{12}^+$ the projection
onto the antisymmetric subspace of $\cH\otimes\cH$ we can
rewrite
\begin{equation*}
\frac{1}{d}I\otimes I-\frac{1}{d_2}P_{12}^+=
\frac{1}{d}P_{12}^-+\frac{d-1}{d(d+1)}P_{12}^+ \, .
\end{equation*}
 Since this is positive full-rank
operator it follows that also $\overline{q}_{jk}(\A=\B)>0$ for all test states.
Therefore, the occurence of different
outcomes cannot be used to unambiguously conclude that the measurements
are different. However, $\overline{q}_{jj}(\A=\B)=0$
if $\Pi_\varrho\leq P_{12}^-$,
hence using a test state supported on the antisymmetric subspace and observing
the same outcomes implies that $\A\neq\B$ with certainty.

In summary, the  identicality
of unknown sharp non-degenerate observables cannot
be unambiguously confirmed if each of the labeled apparatuses is used
only once. Using an antisymmetric test state $\varrho$ and observing
the same outcomes on both apparatuses lead us to unambiguous conclusion
that the apparatuses are different. For fixed observables $\A\neq\B$ the conditional
probability of unambiguous conclusion reads
\be
q_{\rm same}(\A,\B)=\sum_j \tr{\varrho \A_j\otimes\B_j}\, .
\ee
On average
\be
\nonumber
\overline{q}_{\rm same}(\A\neq\B)=d \overline{q}_{jj}(\A\neq\B)=\frac{1}{d}\, .
\ee
This value gives the average conditional success probability
for revealing the difference of the compared
labeled non-degenerate observables. It is independent
of the used test state, however, the unambiguous no-error conditions restricts
the possible test states to so-called antisymmetric states, i.e. those supported
only in the antisymmetric subspace spanned by $P_{12}^-$. Let us stress that
if we choose a test state
$\varrho=\frac{1}{d_-}P^-$, then $q_{\rm same}(\A,\B)>0$ whenever
$\A\neq\B$.

\section{Comparison of unlabeled measurements}\label{sec:unlabeled}

In this section we assume that the outcomes of
measurement devices are not labeled. As previously, our goal
is to design an experiment from which we are able to unambiguously conclude
whether these apparatuses are same or not. But same now means that the observables are equivalent in the unlabeled sense.

Consider a pair of known but unlabeled measurements $\A$ and $\B$.
A single usage of each of the apparatuses leads us to outcome
$j$ on $\A$-apparatus and $a$ on $\B$-apparatus with
probability
\be
p_{j,a}=\frac{1}{n^2}\sum_{j^\prime,a^\prime}\tr{\varrho \A_{j^\prime}\otimes\B_{a^\prime}}=\frac{1}{n^2}
\tr{\varrho}\,.
\ee
Since this probability is independent on whether $\A=\B$ or $\A\neq\B$
none of the outcomes can be used to make a conclusion. In fact $p_{j,a}$ is
independent of particular observables at all. Hence, we need to use the unlabeled
apparatuses more times. In particular, if each of them is used twice, then
\begin{equation}
\nonumber
\begin{array}{rcl}
p_{jk,ab}&=&\frac{1}{n! n!}\sum_{\pi,\pi^\prime}
\tr{\varrho \A_{\pi(j)}\otimes\A_{\pi(k)}\otimes\B_{\pi^\prime(a)}
\otimes\B_{\pi^\prime(b)}}\\
&=& \left\{\begin{array}{lcl}
\frac{1}{d^2}\tr{\varrho \A_{\rm same}\otimes\B_{\rm same}}
& {\rm if}& j=k,a=b\\
\frac{1}{d^2(d-1)}\tr{\varrho \A_{\rm same}\otimes\B_{\rm diff}} \,\ & {\rm if}& j=k,a\neq b\\
\frac{1}{d^2(d-1)}\tr{\varrho \A_{\rm diff}\otimes\B_{\rm same}} \,\ & {\rm if}& j\neq k,a=b\\
\frac{1}{d^2(d-1)^2}\tr{\varrho \A_{\rm diff}\otimes\B_{\rm diff}} & {\rm if}& j\neq k,a\neq b\,,
\end{array}
\right.
\end{array}
\end{equation}
where
\be
\nonumber
\A_{\rm same}=\sum_j \A_j\otimes\A_j\, , \quad
\A_{\rm diff}=\sum_{j\neq k} \A_j\otimes\A_k\,,
\ee
and similarly for $\B_{\rm same}$ and $\B_{\rm diff}$.
We see that irrespectively whether $\A=\B$ or $\A\neq\B$ probability $p_{jk,ab}$ depends only on the mutual relation of the outcomes $j,k$ and $a,b$ of the two usages of the measurement $\A$ respectively $\B$. Hence, it is meaningful to distinguish at most four corresponding classes of outcomes.

For unknown $\A$ and $\B$ ($\A\neq\B$) restricted to be non-degenerate sharp
observables the probability to find the same outcomes on apparatus $\A$
and the same outcomes on apparatus $\B$, respectively, can be expressed
as $p_{\rm same,same}=\tr{\varrho \cO^{\A\neq\B}_{\rm same,same}}$ with
\be
\nonumber
\cO^{\A\neq\B}_{\rm same,same}&=&
d^2\frac{1}{d^2}\int dU dV \A_{\rm same}^U\otimes\B_{\rm same}^V\\
\nonumber
&=&d^2 \int d\psi d\varphi\, \psi\otimes\psi\otimes\varphi\otimes\varphi\\
&=&d^2\overline{\R}_{\rm same}\otimes\overline{\R}_{\rm same}\,,
\ee
where the factor $d^2$ stands
for the number of same outcome pairs that can be observed on individual
apparatuses and we used the definitions
\be
\nonumber
\overline{\R}_{\rm same}&=&\int d\psi \psi\otimes\psi=\frac{1}{d_2}P^+\, ,\\
\nonumber
\overline{\R}_{\rm diff}&=&\int d\psi d\psi_\perp \psi\otimes\psi_\perp=\frac{1}{d}I-\frac{1}{d_2}P^+\,.
\ee
Similarly, for other outcomes we find that
\be
\cO^{\A\neq\B}_{\rm diff,diff}&=&d^2(d-1)^2\overline{\R}_{\rm diff}\otimes\overline{\R}_{\rm diff}\\
\cO^{\A\neq\B}_{\rm diff,same}&=&d^2(d-1)\overline{\R}_{\rm diff}\otimes\overline{\R}_{\rm same}\\
\cO^{\A\neq\B}_{\rm same,diff}&=&d^2(d-1)\overline{\R}_{\rm same}\otimes\overline{\R}_{\rm diff}\, ,
\ee
providing $\A\neq\B$. Let us define operators
\be
& & \Pi_{\rm same,same}^{\A\neq\B} = P_{12}^+\otimes P_{34}^+\,,\\
& & \Pi_{\rm same,diff}^{\A\neq\B} = P_{12}^+\otimes I_{34}\,,\\
& & \Pi_{\rm diff,same}^{\A\neq\B} = I_{12}\otimes P_{34}^+\,,\\
& & \Pi_{\rm diff,diff}^{\A\neq\B} = I_{12}\otimes I_{34}\,,
\ee
that project onto the supports of operators $\cO_{\rm same,same}^{\A\neq\B}$,
$\cO_{\rm same,diff}^{\A\neq\B}$, $\cO_{\rm diff,same}^{\A\neq\B}$,
$\cO_{\rm diff,diff}^{\A\neq\B}$, respectively.

If $\A=\B$, then
\be
\nonumber
& & \cO^{\A=\B}_{\rm same,same}=
d^2\frac{1}{d^2}\int dU \A^U_{\rm same}\otimes\A_{\rm same}^U\\
\nonumber & & =d\int d\psi
\psi\otimes\psi\otimes\psi\otimes\psi
\\ \nonumber & & \qquad
+d(d-1)\int d\psi d\psi_\perp\,
\psi\otimes\psi\otimes\psi_\perp\otimes\psi_\perp\,,
\ee
where, in the second term the integration over $d\psi_\perp$
runs over all vectors orthogonal to a fixed $\psi$. In a general
case the operators $\cO_{x,y}^{\A=\B}=\int dU \A^U_x\otimes\A^U_y$
read
\be
\nonumber & & \cO_{\rm same,diff}^{\A=\B}=
d(d-1)\int
\psi\otimes\psi\otimes[\psi\otimes\psi_\perp+\psi_\perp\otimes\psi]\\
\nonumber & &\qquad\qquad\qquad
 +\frac{d!}{(d-3)!}\int\psi\otimes\psi\otimes\psi^\prime\otimes\psi^\prime_\perp\,, \\
\nonumber & & \cO_{\rm diff,same}^{\A=\B}=d(d-1)\int
\psi\otimes\psi_\perp\otimes[\psi\otimes\psi+\psi_\perp\otimes\psi_\perp]\\
\nonumber & & \qquad\qquad\qquad
+\frac{d!}{(d-3)!}\int\psi^\prime\otimes\psi_\perp^\prime\otimes\psi\otimes\psi\,,\\
\nonumber & & \cO_{\rm diff,diff}^{\A=\B}=d(d-1)\int
\psi\otimes\psi_\perp\otimes[\psi\otimes\psi_\perp+\psi_\perp\otimes\psi]\\
\nonumber & & \qquad\qquad\quad
+\frac{d!}{(d-3)!}\int\psi\otimes\psi_\perp\otimes[\psi\otimes\psi^\prime+\psi^\prime\otimes\psi]\\
\nonumber & & \qquad\qquad\quad
+\frac{d!}{(d-3)!}\int\psi\otimes\psi_\perp\otimes[\psi_\perp\otimes\psi^\prime+\psi^\prime\otimes\psi_\perp]\\
\nonumber & & \qquad\qquad\quad
+\frac{d!}{(d-4)!}\int\psi\otimes\psi_\perp\otimes\psi^\prime\otimes\psi^\prime_\perp\\
\ee
where for simplicity we
do not write explicitly the Haar measures
$d\psi$, $d\psi^\prime$, $d\psi_\perp$, $d\psi_\perp^\prime$ and
$\psi^\prime,\psi^\prime_\perp$ are vectors orthogonal to $\psi$ and $\psi_\perp$.
Of course, $\ip{\psi}{\psi_\perp}=\ip{\psi^\prime}{\psi_\perp^\prime}=0$.
Since for qubits the Hilbert space is two dimensional the terms
containing $\psi^\prime\otimes\psi^\prime_\perp$ do not appear in these
expressions for qubits. There are no two orthogonal vectors to a fixed
$\psi$  in such case.

Let us note that the integration leading to $\cO_{x,y}^{\A\neq\B}$
includes the integration covered in $\cO_{x,y}^{\A=\B}$.
Therefore,
\be
\Pi_{x,y}^{\A=\B}\leq \Pi_{x,y}^{\A\neq\B}\,,
\ee
which implies that whenever
$p_{x,y}(\A\neq\B)=\tr{\varrho \cO_{x,y}^{\A\neq B}}0$,
then also $p_{x,y}(\A=\B)=\tr{\varrho\cO_{x,y}^{\A=\B}}=0$, hence, in two shots
we cannot unambiguously conclude that the apparatuses are the same.
We can only approve the difference of measurement devices.

In what follows we are going to specify for which test states
and for which outcomes $x,y\in\{\textrm{same},\textrm{diff}\}$ the no-error conditions
$\tr{\varrho\cO_{x,y}^{\A=\B}}=0$
are satisfied and simultaneuously,
whether the associated conditional success probability rates
$p_{\rm success}=p_{x,y}=\tr{\varrho\cO_{x,y}^{\A\neq\B}}>0$
are nonvanishing. We shall show that for qubits ($d=2$)
\be
\Pi_{\rm same,same}^{\A\neq\B}&=&\Pi_{\rm same,same}^{\A=\B}+Q_{\rm same,same}\,, \\
\Pi_{\rm same,diff}^{\A\neq\B}&=& \Pi_{\rm same,diff}^{\A=\B}+Q_{\rm same, diff}\,,\\
\Pi_{\rm diff,same}^{\A\neq\B}&=& \Pi_{\rm diff,same}^{\A=\B}+Q_{\rm diff,same}\,,\\
\Pi_{\rm diff,diff}^{\A\neq\B}&=& \Pi_{\rm diff,diff}^{\A=\B}+Q_{\rm diff,diff}\,,
\ee
where $Q_{\rm same,same}=O$, $Q_{\rm diff,diff}\neq Q_{\rm same,diff}=Q_{\rm diff,same}$ are
projections forming the relevant parts of the supports of potential
test states $\varrho$ enabling us to conclude the difference. That is,
we shall see that three out of four outcomes can be used to make
the unambiguous conclusion.

\subsection{$\cO_{\rm same,same}$}\label{sec:same_same}
Evaluating the operator $\cO_{\rm same,same}^{\A=\B}$ we obtain
\be
\nonumber
& & \frac{1}{d(d-1)}\cO_{\rm same,same}^{\A=\B}=\int \psi^{\otimes 4}+
(d-1)\int\psi^{\otimes 2}\otimes\psi_\perp^{\otimes 2}\\
& & = \frac{1}{d_4}P_{1234}^++\frac{2(d-1)}{d(d-1)}R_{12-34} P_{34}^+\, ,
\label{intsamesame}
\ee
where
\be
\nonumber
R_{12-34}&=&
\int\psi^{\otimes 2}
\otimes(I-\psi)^{\otimes 2}\\
\nonumber &=&
\frac{1}{d_2}P_{12}^++\frac{1}{d_4}P_{1234}^+
-\frac{1}{d_3}(P_{123}^++P_{124}^+)\, .
\ee
Due to positivity of operators in Eq. (\ref{intsamesame})
the unambiguous no-error conditions require that
\be
\nonumber
\tr{\varrho P_{1234}^+}=0\,,\quad
\tr{\varrho R_{12-34} P_{34}^+}=0\,,
\ee
hold simultaneously. Hence, the support of $R_{12-34} P_{34}^+$
is of interest for us and in particular we should decide
whether it is different from
$\Pi_{\rm same,same}^{\A\neq\B}=P_{12}^+\otimes P_{34}^+$. If yes, then
we can use this outcome for making the unambiguous conclusion.

Let us analyze properties of $R_{12-34}$ and its terms. First of all by
definition $R_{12-34} P_{34}^+$ is a positive operator, hence
necessarily $[R_{12-34},P_{34}^+]=0$ and also
$[P_{123}^++P_{124}^+,P_{34}^+]=0$. The support of the projections $P_{12}^+$,
$P_{1234}^+$, $P_{123}^+$, and $P_{124}^+$ contains the completely symmetric
subspace spanned by $P_{1234}^+$. As it is shown in Appendix
it is their greatest joint subspace and since
$\frac{1}{d_2}+\frac{1}{d_4}-\frac{2}{d_3}>0$ the operator
$R_{12-34}$ is indeed supported on the whole $P_{1234}^+$.

It remains to analyze the properties of $R_{12-34} P_{34}^+$ on the
subspace $Q_{12}^+=P_{12}^+\otimes P_{34}^+-P_{1234}^+$.
In particular, we are interested whether
\be
\bra{\varphi}\frac{1}{d_2}Q_{12}^+ - \frac{1}{d_3}(Q_{123}+Q_{124})
\ket{\varphi}>0
\ee
for all $\ket{\varphi}$ from the support of $Q_{12}^+$, where
$Q_{123}=P_{123}^+-P_{1234}^+$, $Q_{124}=P_{124}^+-P_{1234}^+$.
For qubits these subspaces are described in details
in Appendix \ref{sec:subspaces}, where it is shown that
the operator $Q_{123}+Q_{124}$ have two nonzero eigenvalues
$4/3$ and $2/3$. However, the eigenvectors associated with
$4/3$ are from the subspace spanned by $P_{12}^+\otimes P_{34}^-$,
which is irrelevant due to multiplication of $R_{12-34}$ by $P_{34}^+$. The eigenvectors associated with the
eigenvalue $2/3$ are from $P_{12}^+\otimes P_{34}^+$, thus
$\bra{\varphi}Q_{123}+Q_{124}\ket{\varphi}\leq 2/3$ for all
$\ket{\varphi}\in P_{12}^+\otimes P_{34}^+\ge Q_{12}^+$.
Since $d_2=3$, $d_3=4$
\be
\bra{\varphi}\frac{1}{3}Q_{12}^+ - \frac{1}{4}(Q_{123}+Q_{124})
\ket{\varphi}\ge \frac{1}{3}-\frac{1}{6}>0\, .
\ee
As a result we have shown that support of
$R_{12-34}P_{34}^+$ equals to support of $P_{12}^+\otimes P_{34}^+$, thus
$\Pi_{\rm same,same}^{\A=\B}=P_{12}^+\otimes P_{34}^+
=\Pi_{\rm same,same}^{\A\neq\B}$. In summary, an observation of pairs of
same outcomes on both apparatuses cannot be used to make any unambiguous
conclusion, because $Q_{\rm same,same}=O$.

\subsection{$\cO_{\rm diff,diff}$}{\label{sec:diff_diff}}

In this case our aim is to show that $Q_{\rm diff,diff}\neq O$.
For qubits there are at most two mutually orthogonal vectors, hence
\be
\nonumber
\cO_{\rm diff,diff}^{\A=\B}=d(d-1)
\int \psi\otimes\psi_\perp\otimes(\psi\otimes\psi_\perp+\psi_\perp\otimes\psi)\,.
\ee
Let us remind that for larger systems, this expression contains
additional terms. Using the operators $R_{13-24}$, $R_{14-23}$ introduced
in a similar way as $R_{12-34}$ defined in the previous section we obtain
\be
\cO_{\rm diff, diff}^{\A=\B}=2(R_{13-24}P_{24}^++R_{14-23}P_{23}^+).
\ee
Using the same arguments as for $R_{12-34}$ we find that
$R_{13-24}P_{24}^+$ is supported on $P_{13}^+\otimes P_{24}^+$
and $R_{14-23}P_{23}^+$ is supported on $P_{14}^+\otimes P_{23}^+$.
Therefore, for the test state $\varrho$ we can write the following
no-error condition
\be
0=\tr{\varrho (P_{13}^+\otimes P_{24}^++P_{14}^+\otimes P_{23}^+)}\, .
\ee
The completely symmetric subspace $P_{1234}^+$
is the greatest joint subspace of $P_{13}^+\otimes P_{24}^+$
and $P_{14}^+\otimes P_{23}^+$. According to Appendix \ref{sec:sub_1}
the support of $P_{13}^+\otimes P_{24}^++P_{14}^+\otimes P_{23}^+$
is 13 dimensional, because $d_4=5$ and
$Q_{13}^+=P_{13}^+\otimes P_{24}^+-P_{1234}^+$
and $Q_{14}^+=P_{14}^+\otimes P_{23}^+-P_{1234}^+$ are both four dimensional.
Since the total Hilbert space $\cH^{\otimes 4}$ for qubits is 16-dimensional,
it follows that test states satisfying the no-error conditions
live in a three-dimensional subspace. In Appendix \ref{sec:sub_1}
it is shown that this subspace is a linear span of vectors
\be
\nonumber
\ket{\kappa_1}&=&\frac{1}{\sqrt{2}}
(\ket{00}\ket{\psi^+}-\ket{\psi^+}\ket{00})\,,
\\ \nonumber\ket{\kappa_2}&=&\frac{1}{\sqrt{2}}(\ket{0011}-\ket{1100})\,,
\\ \nonumber\ket{\kappa_3}&=&\frac{1}{\sqrt{2}}
(\ket{11}\ket{\psi^+}-\ket{\psi^+}\ket{11})\, ,
\ee
where $\ket{\psi^+}=\frac{1}{\sqrt{2}}(\ket{01}+\ket{10})$. Thus,
$Q_{\rm diff,diff}=\sum_j \ket{\kappa_j}\bra{\kappa_j}
\leq Q_{12}^+\leq P_{12}^+\otimes P_{34}^+$ and arbitrary
test state $\varrho\leq Q_{\rm diff,diff}$
satisfies the no-error condition.

Let us optimize the conditional probability
\be
\overline{p}_{\rm diff,diff}(\A\neq\B)=
\tr{\varrho \cO_{\rm diff,diff}^{\A\neq\B}}\,
\ee
where
\be
\nonumber
\cO_{\rm diff,diff}^{\A\neq\B}&=&4(\frac{1}{2}I-\frac{1}{3}P_{12}^+)
\otimes(\frac{1}{2}I-\frac{1}{3}P_{34}^+)\\
\nonumber &=&
I-\frac{2}{3}(P_{12}^++P_{34}^+)+\frac{4}{9}P_{12}^+\otimes P_{34}^+\, .
\ee
Arbitrary pure state $\ket{\varphi}\in Q_{\rm diff,diff}$ is an eigenvector
of projections $P_{12}^+$, $P_{34}^+$ and $P_{12}^+\otimes P_{34}^+$.
Therefore, the probability is independent of the test states
$\varrho\leq Q_{\rm diff,diff}$ and reads
\be
\overline{p}_{\rm diff,diff}(\A\neq\B)=1-\frac{4}{3}+\frac{4}{9}=\frac{1}{9}\,.
\ee

\subsection{$\cO_{\rm same,diff}$}{\label{sec:same_diff}}
For qubits
\be
\nonumber
\cO_{\rm same,diff}^{\A=\B}&=&
d(d-1)\int \psi^{\otimes 2}\otimes(\psi\otimes\psi_\perp+\psi_\perp\otimes\psi)\\
\nonumber &=&
d\left(\frac{1}{d_3}(P_{123}^++P_{124}^+)-\frac{2}{d_4}P_{1234}^+\right)\,,
\ee
and since $P_{1234}^+ \leq P_{123}^+,P_{124}^+$; $1/d_3>1/d_4$ we can conclude that no-error unambiguous
condition reads
\be
\tr{\varrho (P_{123}^++P_{124}^+)}=0\, .
\ee
Let us remind that $\Pi_{\rm same,diff}^{\A\neq\B}=P_{12}^+$
and $P_{123}^+,P_{124}^+\leq P_{12}^+$. The question is whether
$\Pi_{\rm same,diff}^{\A=\B}=P_{12}^+$, or not. We know
(see Appendix \ref{sec:subspaces}) that
$P_{123}^+,P_{124}^+$ are not orthogonal, however, their greatest joint subspace
is the completely symmetric one. The dimension of $P_{12}^+$ is 12, whereas
the total support of $P_{123}^++P_{124}^+$ is 11 dimensional. It follows that
there exist a unique vector such that
$\Pi_{\rm same,diff}^{\A\neq\B}\ket{\varphi_Q}=\ket{\varphi_Q}$,
and, simultaneuously, $\Pi_{\rm same,diff}^{\A=\B}\ket{\varphi_Q}=0$, thus,
$Q_{\rm same,diff}=\ket{\varphi_Q}\bra{\varphi_Q}$. For such test state the
observation of this outcome leads to unambiguous confirmation of the
difference of the measurement devices.

\subsection{$\cO_{\rm diff,same}$}{\label{sec:diff_same}}

There is no substantial difference in the analysis of this
case and the previous one. We only need to exchange the role
of pairs of indexes $12$ and $34$. Therefore, there exists a unique
vector $\ket{\varphi_Q^\prime}$ such that
$\Pi_{\rm diff,same}^{\A=\B}\ket{\varphi_Q^\prime}=0$,
but $\Pi_{\rm diff,same}^{\A\neq\B}\ket{\varphi_Q^\prime}=
P_{34}^+\ket{\varphi^\prime_Q}=\ket{\varphi_Q^\prime}$.
Surprisingly, we shall see that $\ket{\varphi_Q^\prime}\equiv\ket{\varphi_Q}$,
which means that the same test state $\ket{\varrho_Q}$ guarantees
the unambiguity of both outcomes $\cO_{\rm same,diff},\cO_{\rm diff,same}$.

On the systems $j$ and $k$ we define a singlet vector as $\ket{\psi^-_{jk}}
=\frac{1}{\sqrt{2}}(\ket{01}_{jk}-\ket{10}_{jk})$.
After a short calculation one can verify that the vector
\be
\ket{\varphi_Q}=\frac{1}{\sqrt{3}}
(\ket{\psi^-_{13}\otimes\psi_{24}^-}
+\ket{\psi_{14}^-\otimes\psi^-_{23}})
\ee
satisfies all the required properties, i.e. it is symmetric
with respect to $1\leftrightarrow 2$, $3\leftrightarrow 4$ exchanges,
i.e. $\Pi^{\A\neq\B}_{\rm same,diff}\ket{\varphi_Q}
=\Pi_{\rm diff,same}^{\A\neq\B}\ket{\varphi_Q}=\ket{\varphi_Q}$,
and $P_{123}^+\ket{\varphi_Q}=P_{124}^+\ket{\varphi_Q}=P_{134}^+\ket{\varphi_Q}
=P_{234}^+\ket{\varphi_Q}=0$, because both terms of
$\ket{\varphi_Q}$ are antisymmetric exactly in one pair of all considered
triples of indexes.

Using $\ket{\varphi_Q}$ as the test state we get
\be
\nonumber
& &
\overline{p}_{\rm same,diff}(\A\neq\B)=\bra{\varphi_Q}\cO_{\rm same,diff}^{\A\neq\B}
\ket{\varphi_Q}\\
\nonumber & & \quad =
\frac{4}{3}\bra{\varphi_Q}
\frac{1}{6}P_{12}^+\otimes P_{34}^++\frac{1}{2}P_{12}^+\otimes P_{34}^-
\ket{\varphi_Q}\, .
\ee
Similarly, we find
\be
\nonumber
p_{\rm diff,same}(\A\neq\B)=
\frac{4}{3}\bra{\varphi_Q}
\frac{1}{6}P_{12}^+\otimes P_{34}^++\frac{1}{2}P_{12}^-\otimes P_{34}^+
\ket{\varphi_Q}\, .
\ee
Since $P_{12}^+\ket{\varphi_Q}=P_{34}^+\ket{\varphi_Q}=\ket{\varphi_Q}$
implies $P_{12}^+\otimes P_{34}^+\ket{\varphi_Q}=\ket{\varphi_Q}$ and
\be
\nonumber
\bra{\varphi_Q} P_{12}^-\otimes P_{34}^+\ket{\varphi_Q}=
\bra{\varphi_Q} P_{12}^+\otimes P_{34}^-\ket{\varphi_Q}=0\, ,
\ee
we obtain
\be
\overline{p}_{\rm same,diff}(\A\neq\B)
+\overline{p}_{\rm diff,same}(\A\neq\B)=\frac{4}{9}\, .
\ee
This gives a better success rate than we achieved for the outcome
$\cO_{\rm diff,diff}$. Unfortunately, $\ket{\varphi_Q}\not\in Q_{\rm diff,diff}$.
In conclusion, $\overline{p}=4/9$ is the optimal value of the
average success rate for unambiguous comparison of unlabeled
qubit non-degenerate sharp observables in two shots.

Consider a pair of observables $\A=\{\psi,\psi_\perp\}$,
$\B=\{\varphi,\varphi_\perp\}$ such that $\psi\neq\varphi$.
Then the projections
\be
\nonumber
\cO_{\rm diff,same}&=&(\psi\otimes\psi_\perp+\psi_\perp\otimes\psi)\otimes (\varphi\otimes\varphi+
\varphi_\perp\otimes\varphi_\perp)\,, \\
\nonumber
\cO_{\rm same,diff}&=&(\psi\otimes\psi+
\psi_\perp\otimes\psi_\perp)\otimes
(\varphi\otimes\varphi_\perp+\varphi_\perp\otimes\varphi)\,.
\ee
The success probability of revealing their difference
using the test state $\ket{\varphi_Q}$ reads
\be
p_{\rm success}(\psi,\varphi)=
\bra{\varphi_Q}\cO_{\rm same,diff}+\cO_{\rm diff,same}\ket{\varphi_Q}\,.
\ee
Let us note that in a fixed orthonormal basis $\ket{\psi},\ket{\psi_\perp}$
the test state $\ket{\varphi_Q}$ takes the form
\be
\nonumber
\ket{\varphi_Q}=\frac{1}{\sqrt{3}}\left(
\ket{\psi^{\otimes 2}\otimes\psi_\perp^{\otimes 2}}
+\ket{\psi_\perp^{\otimes 2}\otimes\psi^{\otimes 2}}
-\ket{\psi^+\otimes\psi^+}
\right)\,,
\ee
where $\ket{\psi^+}=\frac{1}{\sqrt{2}}(\ket{\psi\otimes\psi_\perp}+
\ket{\psi_\perp\otimes\psi})$. Using the identities
$|\ip{\psi}{\varphi}|=|\ip{\psi_\perp}{\varphi_\perp}|=\cos\theta$,
$|\ip{\psi}{\varphi_\perp}|=|\ip{\psi_\perp}{\varphi}|=\sin\theta$
a direct calculation gives
\be
\nonumber
\bra{\varphi_Q}\cO_{\rm same,diff}\ket{\varphi_Q}
&=&\frac{1}{3}\bra{\psi_\perp^{\otimes 2}}\varphi\otimes\varphi_\perp+\varphi_\perp\otimes\varphi\ket{\psi_\perp^{\otimes 2}}
\\ \nonumber & & +
\frac{1}{3}\bra{\psi^{\otimes 2}}\varphi\otimes\varphi_\perp+\varphi_\perp\otimes\varphi\ket{\psi^{\otimes 2}}\\
\nonumber &=&
\frac{4}{3}|\ip{\psi}{\varphi}|^2|\ip{\psi_\perp}{\varphi}|^2\,.
\ee
Since $\bra{\varphi_Q}\cO_{\rm same,diff}\ket{\varphi_Q}
=\bra{\varphi_Q}\cO_{\rm diff,same}\ket{\varphi_Q}$ the success
probability reads
\be
p_{\rm success}(\psi,\varphi)=\frac{2}{3}(\sin{2\theta})^2\,.
\ee
It vanishes only if $\theta=0$, or $\theta=\pi/2$, hence
$\psi\equiv\varphi$, or $\psi\equiv\varphi_\perp$, respectively.
As a result we get that the optimal test state detects unambiguously
the difference for any pair of non-equivalent sharp qubit observables with
strictly nonzero success probability. The actual probability depends on
the angle between the observables. In fact,
if sharp qubit POVMs are understood as ideal Stern-Gerlach
apparatuses, then $\alpha=2\theta$ is the angle between the
measured spin directions. The probability achieves its maximum for
orthogonal spin directions as one would expect.

\section{Summary}\label{sec:summary}

We have investigated the problem of unambiguous comparison of
quantum measurements. We restricted our analysis
to subset of sharp non-degenerate observables
that can be associated with non-degenerate selfadjoint operators.
Let us note that without any restriction the comparison problem has only a trivial
solution.

We distinguished two different types of measurement apparatuses
depending whether the labels of their outcomes are apriori given,
or not. We give solution to single shot comparison of labeled measurements
in arbitrary dimension. For unlabeled measurements the single usage of
each of the apparatuses is not sufficient. In the two shots scenario
we find solution for unlabeled qubit measurement apparatuses. In both cases,
the unambiguous confirmation of the equivalence of
measurements is not possible. Similarly, as in the case of pure
states and unitary channels, also for sharp non-degenerate observables
only the difference can be unambiguously concluded.

In summary, for the measurement apparatuses with labeled outcomes
the optimal procedure exploits the so-called antisymmetric test states.
For any such test state $\varrho$ the success is associated
with the observation of the same outcomes. The difference of observables
can be concluded with the average conditional probability
\be
\overline{q}_{\rm success}(\A\neq\B)=1/d\, .
\ee

\begin{figure}
\includegraphics[width=7cm]{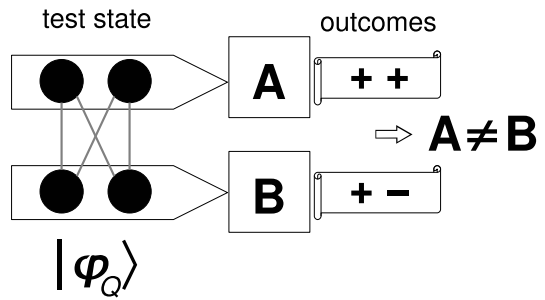}
\caption{Illustration of the optimal scheme for unambiguous
comparison of qubit apparatuses leading to unambiguous
conclusion $A\neq B$ with average conditional probability $4/9$.}
\label{fig:comp_scheme}
\end{figure}

In the case of unlabeled measurements individual outcomes can be
associated with
an unambiguous conclusion only if the
support of the test
state belongs to at least one of the subspaces spanned by projections
$1-\Pi_{x,y}^{\A=\B}$, $x,y\in \{{\rm same,diff}\}$.
We showed that only part of the test state acting
on the support of the projections
$Q_{\rm same,same}=O$, $Q_{\rm diff,diff}$ and
$Q_{\rm same,diff}=Q_{\rm diff,same}=\ket{\varphi_Q}\bra{\varphi_Q}$
may contribute to the success probability. Out of these possibilities, it turns out that the optimal
test state is
\be
\ket{\varphi_Q}=\frac{1}{\sqrt{3}}(\ket{\psi^-_{13}\otimes\psi_{24}^-}
+\ket{\psi_{14}^-\otimes\psi^-_{23}}) \,,
\ee
for which the average conditional probability of the
unambiguous conclusion equals
\be
\overline{p}_{\rm success}(\A\neq\B)=4/9\, .
\ee
Using such test state and finding
on one of the measurement different outcomes, whereas on the
second the same outcomes, we can conclude with certainty
that the apparatuses are different. This scheme is illustrated
on Fig.~\ref{fig:comp_scheme}.

Let us compare these success probabilities with the comparison problem for pure states
and unitary channels. In particular, for single shot comparisons
\be
\overline{p}_{\rm state}&=& (d-1)/2d\,,\\
\overline{p}_{\rm unitary} &=& (d+1)/2d\,.
\ee
We see that unlike for states and channels the success rate
for comparison of labeled measurements vanishes as the dimension
is increasing. Unfortunately, for unlabeled measurements on systems
of larger dimensions the situation is more complex and
two shots are not sufficient to make any unambiguous conclusion.
The problem is still open and will be analyzed elsewhere.

\appendix
\section{Subspaces}{\label{sec:subspaces}}
In this appendix we shall analyze the subspaces of four quantum systems
$\cH^{\otimes 4}$, especially four qubits. Let us start with the simpler
case of $\cH\otimes\cH$. Denote by $\ket{j}$ the basis of $\cH$ and define
\be
\ket{\varphi^\pm_{jk}}=\frac{1}{\sqrt{2}}(\ket{j\otimes k}
\pm\ket{k\otimes j})\, .
\ee
for $j<k$. For $j=k$
\be
\ket{\varphi^+_{jj}}=\ket{j\otimes j}\, .
\ee
These vectors form an orthonormal bases of symmetric and antisymmetric
subspaces of $\cH\otimes\cH$, i.e. they define the projections
$P_{12}^\pm =\sum_{j\leq k} \ket{\varphi_{jk}^\pm}\bra{\varphi_{jk}^\pm}$.

We shall use the notation  $P_{12}^\pm=P_{12}^\pm\otimes I_{34}
=P_{12}^\pm\otimes (P_{34}^++P_{34}^-)$.
Let us stress that $P_{1234}^+\leq P_{123}^+\leq P_{12}^+$.
We shall be interested in properties of projections that are substracted
from other projections to create the projections onto the
completely symmetric subspace, for example,
operators $Q_{12}=P_{12}^+-P_{1234}^+$ and $Q_{123}=P_{123}^+-P_{1234}^+$.
Similar notations, definitions and relations hold also for other
combination of indexes.

For qubits ${\rm dim} P_{12}^+=d^2 \cdot d_2=12$,
${\rm dim}P_{12}^+\otimes P_{34}^+=d_2^2=9$,
${\rm dim} P_{123}^+={\rm dim} P_{124}^+=d\cdot d_3=8$ and
${\rm dim} P_{1234}^+=d_4=5$, thus,
${\rm dim}Q_{123}={\rm dim}Q_{124}=3$ and $Q_{12}=7$., etc.

\subsection[A.1]{$P_{12}^+\otimes P_{34}^+$ and $P_{1234}^+$}\label{sec:sub_1}
Let us start with the analysis of the subspace of $P_{12}^+$
not contained in $P_{1234}^+$, i.e. with $Q_{12}$. In the first step, let us
split $Q_{12}$ into  $Q_{12}=Q_{12}^-+Q_{12}^+$, where $Q_{12}^\pm =
P_{12}^+\otimes P_{34}^\pm - P_{1234}^+$, $Q_{12}^-=P_{12}^+\otimes P_{34}^-$. Due to asymmetry
of $P_{12}^+\otimes P_{34}^-$ in $3\leftrightarrow 4$ exchange
the projections $P_{1234}^+$ {\bf and $Q_{12}^-$} are orthogonal. For $Q_{12}^+$ the situation is more
tricky. Our goal is to design a basis of the support of $Q_{12}^+$.
The completely symmetric subspace $P_{1234}^+$ is spanned by the following
orthonormal basis
\be
\nonumber
\ket{\eta_0}&=&\ket{\varphi^+_{00}\otimes\varphi_{00}^+}\\
\nonumber
\ket{\eta_1}&=&\frac{1}{\sqrt{2}}(\ket{\varphi^+_{00}\otimes\varphi_{01}^+}+
\ket{\varphi^+_{01}\otimes\varphi_{00}^+})\\
\nonumber
\ket{\eta_2}&=&\sqrt{\frac{2}{3}}\ket{\varphi_{01}^+\otimes\varphi_{01}^+}+
\sqrt{\frac{1}{6}}(\ket{\varphi_{00}^+\otimes\varphi_{11}^+}+
\ket{\varphi_{11}^+\otimes\varphi_{00}^+})\\
\nonumber
\ket{\eta_3}&=&\frac{1}{\sqrt{2}}(\ket{\varphi^+_{11}\otimes\varphi_{01}^+}+
\ket{\varphi^+_{01}\otimes\varphi_{11}^+})\\
\nonumber
\ket{\eta_4}&=&\ket{\varphi^+_{11}\otimes\varphi_{11}^+}\, .
\ee
Our aim is to specify a basis spanning the support of $Q_{12}^+$. Since
${\rm dim}P_{1234}^+=5$ and ${\rm dim}P_{12}^+\otimes P_{34}^+=9$
it follows we need to find four mutually orthogonal vectors in
$P_{12}^+\otimes P_{34}^+$ that are also orthogonal to vectors $\ket{\eta_j}$.
It is straightforward to verify that the following vectors
\be
\nonumber
\ket{\kappa_1}&=&\frac{1}{\sqrt{2}}
(\ket{\varphi_{00}^+\otimes\varphi_{01}^+}-\ket{\varphi_{01}^+\otimes\varphi_{00}^+})
\\ \nonumber\ket{\kappa_2}&=&\frac{1}{\sqrt{2}}(\ket{\varphi^+_{00}\otimes\varphi^+_{11}}-\ket{\varphi^+_{11}\otimes\varphi^+_{00}}
)
\\ \nonumber\ket{\kappa_2^\prime}&=&
\sqrt{\frac{1}{3}}(\ket{\varphi_{01}^+\otimes\varphi_{01}^+}
-\ket{\varphi^+_{00}\otimes\varphi^+_{11}}
-\ket{\varphi^+_{11}\otimes\varphi^+_{00}})
\\ \nonumber\ket{\kappa_3}&=&\frac{1}{\sqrt{2}}
(\ket{\varphi_{11}^+\otimes\varphi_{01}^+}-\ket{\varphi_{01}^+\otimes\varphi_{11}^+}
)
\ee
form such a basis.

Let us define a swap operator $S_{ab}=P_{ab}^+-P_{ab}^-$
implementing the exchange of the subsystems $a,b$.
This operation is unitary and arbitrary permutation
can be written as a composition of swap operations.
The following identities hold
\be
\nonumber P_{13}^+\otimes P_{24}^+ = S_{23} (P_{12}^+\otimes P_{34}^+) S_{23}\,,\\
\nonumber P_{14}^+\otimes P_{23}^+ = S_{34} (P_{13}^+\otimes P_{24}^+) S_{34}\,,\\
\nonumber P_{12}^+\otimes P_{34}^+ = S_{24} (P_{14}^+\otimes P_{23}^+) S_{24}\,.
\ee
The vectors $\ket{\kappa_1},\ket{\kappa_2},\ket{\kappa_3}$ defined with
respect to division $P_{12}^+\otimes P_{34}^+$ are orthogonal to all
vectors $\ket{\kappa_j},\ket{\kappa_2^\prime}$
defined with respect to splittings
$P_{13}^+\otimes P_{24}^+$ and $P_{14}^+\otimes P_{23}^+$, i.e.
$P_{13}^+\otimes P_{24}^+\ket{\kappa_j}=P_{14}^+\otimes P_{23}^+\ket{\kappa_j}=0$.
However,
$\bra{\kappa_2^\prime}P_{13}^+\otimes P_{24}^+\ket{\kappa_2^\prime}=
\bra{\kappa_2^\prime}P_{14}^+\otimes P_{23}^+\ket{\kappa_2^\prime}=1/4$,
because the vectors $\ket{\kappa_2^\prime}$ defined with
respect to different splittings are mutually nonorthogonal. This means
that the 4 dimensional projections
$Q_{12}^+, Q_{13}^+,Q_{14}^+$ are not orthogonal, however,
there is a three-dimensional subspace of $Q_{12}^+$ (spanned by
vectors $\ket{\kappa_j}$) orthogonal
to both $Q_{13}^+$ and $Q_{14}^+$.

\subsection[A.2]{$P_{123}^++P_{124}^+$}\label{sec:sub_2}
For the purposes of this paper it is of interest
to analyze the relation of the supports of projections
$P_{123}^+$ and $P_{124}^+$.
The swap operator $S_{34}$ can be written as
a composition $S_{34}=S_{24} S_{23} S_{24}$.
Consider a vector $\ket{\varphi}$ belonging
to both subspaces, i.e. $P_{123}^+\varphi=P_{124}^+\ket{\varphi}=\ket{\varphi}$.
For such vector $S_{12}\ket{\varphi}=S_{13}\ket{\varphi}=
S_{14}\ket{\varphi}=S_{23}\ket{\varphi}=S_{24}\ket{\varphi}=\ket{\varphi}$
and therefore also $S_{34}\ket{\varphi}=S_{24}S_{23}S_{24}
\ket{\varphi}=\ket{\varphi}$, hence the state $\varphi$
is symmetric also with respect to
exchange $3\leftrightarrow 4$. Consequently, it is invariant under the swap
of arbitrary subsystems, i.e. it belongs to the
completely symmetric subspace. Therefore, the greatest joint
subspace of supports of $P_{123}^+$ and $P_{124}^+$ corresponds to
the projection $P_{1234}^+$.

Further we shall prove that the projections
$Q_{123}=P_{123}^+-P_{1234}^+$ and $Q_{124}=P_{124}^+-P_{1234}^+$
are not mutually orthogonal and we shall specify
the support of $Q_{123}+Q_{124}$. It is relatively
stragihtforward to verify that the following unnormalized vectors
\be
\nonumber
\ket{\omega_1}&=&
\ket{\varphi_{00}^+}_{12}\ket{\varphi_{01}^-}_{34}
+\ket{\varphi_{00}^+}_{13}\ket{\varphi_{01}^-}_{24}
+\ket{\varphi_{00}^+}_{23}\ket{\varphi_{01}^-}_{14}\,,\\
\nonumber
\ket{\omega_2}&=&
\ket{\varphi_{00}^+\otimes\varphi_{11}^+}
-\ket{\varphi_{11}^+\otimes\varphi_{00}^+}
+2\ket{\varphi_{01}^+\otimes\varphi_{01}^-}\,,
\\ \nonumber
\ket{\omega_3}&=&
\ket{\varphi_{11}^+}_{12}\ket{\varphi_{01}^-}_{34}
+\ket{\varphi_{11}^+}_{13}\ket{\varphi_{01}^-}_{24}
+\ket{\varphi_{11}^+}_{23}\ket{\varphi_{01}^-}_{14}\, ,
\ee
form an orthogonal basis of the support of $Q_{123}$.
These vectors are orthogonal to vectors $\ket{\eta_j}$ forming
the completely symmetric subspace. In fact, they are completely
symmetric only with respect to three indexes ($123$), but they
not with respect to exchanges with the fourth qubit, hence,
$P_{12}^+\otimes P_{34}^+\ket{\omega_j}$ is not proportional to
$\ket{\omega_j}$. In the same way we can design a basis for each
$Q_{jkl}$, in particular, for $Q_{124}$
\be
\nonumber
\ket{\omega_1^\prime}&=&
-\ket{\varphi_{00}^+}_{12}\ket{\varphi_{01}^-}_{34}
+\ket{\varphi_{00}^+}_{14}\ket{\varphi_{01}^-}_{23}
+\ket{\varphi_{00}^+}_{24}\ket{\varphi_{01}^-}_{13}\,,\\
\nonumber
\ket{\omega_2^\prime}&=&
\ket{\varphi_{00}^+\otimes\varphi_{11}^+}
-\ket{\varphi_{11}^+\otimes\varphi_{00}^+}
-2\ket{\varphi_{01}^+\otimes\varphi_{01}^-}\,,
\\ \nonumber
\ket{\omega_3^\prime}&=&
-\ket{\varphi_{11}^+}_{12}\ket{\varphi_{01}^-}_{34}
+\ket{\varphi_{11}^+}_{14}\ket{\varphi_{01}^-}_{23}
+\ket{\varphi_{11}^+}_{24}\ket{\varphi_{01}^-}_{13}\, .
\ee
Since $\ip{\omega_j}{\omega_k^\prime}=-2\delta_{jk}$
the pair of unnormalized vectors $\ket{\omega_j},\ket{\omega_j^\prime}$
forms a two-dimensional subspace orthogonal to remaining
vectors. Equal superpositions
$\ket{\omega_j^+}=\ket{\omega_j}+\ket{\omega_j^\prime}$
are already symmetric in $3\leftrightarrow 4$ exchange, hence
$\ket{\omega_j^+}\in P_{12}^+\otimes P_{34}^+$. On the other hand,
the vectors $\ket{\omega_j^-}=\ket{\omega_j}-\ket{\omega_j^\prime}$
are antisymmetric in $3\leftrightarrow 4$, hence
$\ket{\omega_j^-}\in P_{12}^+\otimes P_{34}^-$. It is
easy to verify that they are orthogonal, i.e.
$\ip{\omega_j^+}{\omega_j^-}=0$, because
$\ip{\omega_j}{\omega_j}=\ip{\omega_j^\prime}{\omega_j^\prime}=6$
and $\ip{\omega_j}{\omega_j^\prime}=\ip{\omega_j^\prime}{\omega_j}=-2$.
Moreover, $\ip{\omega_j^+}{\omega_{j}^+}=8$ and
$\ip{\omega_j^-}{\omega_{j}^-}=16$.
Since $\ket{\omega_j}=\frac{1}{2}(\ket{\omega_j^+}+\ket{\omega_j^-})$,
$\ket{\omega_j^\prime}=\frac{1}{2}(\ket{\omega_j^+}-\ket{\omega_j^-})$
we have
\be
\nonumber
Q_{123}+Q_{124}&=&
\frac{1}{6}\sum_j
(\ket{\omega_j}\bra{\omega_j}+\ket{\omega_j^\prime}\bra{\omega_j^\prime})\\
\nonumber &=&\sum_j\frac{1}{12}
(\ket{\omega_j^+}\bra{\omega_j^+}+\ket{\omega_j^-}\bra{\omega_j^-})\\
\nonumber & =& \sum_j \left(
\frac{4}{3}\frac{1}{16}\ket{\omega^-_j}\bra{\omega^-_j}+
\frac{2}{3}\frac{1}{8}\ket{\omega^+_j}\bra{\omega^+_j}\right)\, ,
\ee
where $\frac{1}{16}\ket{\omega^-_j}\bra{\omega^-_j}$ and
$\frac{1}{8}\ket{\omega^+_j}\bra{\omega^+_j}$ are one-dimensional
projections, hence, we get the spectral decomposition of $Q_{123}+Q_{124}$
with eigenvalues $2/3,4/3$. For our purposes the relevant part
is associated with vectors $\ket{\omega_j^+}$, because $\ket{\omega_j^-}$
are not from the support of $P_{12}^+\otimes P_{34}^+$.

\section*{Acknowledgements}
Authors wish to thank Vlado Bu\v{z}ek and Sergej Nikolajevi\v{c} Filippov for inspiring discussions.
M.Z. and M.S. acknowledge financial support via the European Union projects QAP
2004-IST-FETPI-15848, HIP FP7-ICT-2007-C-221889, and via the projects
APVV-0673-07 QIAM, OP CE QUTE ITMS NFP 262401022,
and CE-SAS  QUTE. M.Z. acknowledges also support of GA\v CR via
project GA201/07/0603. T.H. acknowledges financial support from QUANTOP and Academy of Finland.

\end{document}